# Phenomenology of aftershocks: Once more on the Omori Law

A. V. Guglielmi[1,]*, B. I. Klain[2,]**, A. D. Zavyalov[1,]***, and O. D. Zotov[1, 2,]****

[1]*Institute of Physics of the Earth RAS, Bol'shaya Gruzinskaya str., 10, bld. 1, Moscow, 123242 Russia*

[2]*Borok Geophysical Observatory of Institute of Physics of the Earth, RAS, Borok settlement, Nekouz district, Yaroslavl Region, 152742 Russia*

*e-mail: guglielmi@mail.ru
**e-mail: klb314@mail.ru
***e-mail: zavyalov@ifz.ru
****e-mail: ozotov@inbox.ru

Omori Law describes the evolution of the aftershocks of a strong earthquake. Established at the end of the century before last, it is characterized by the beauty of its form, quite definite clarity, as a result of which it still attracts considerable attention of the geophysical community. In recent years, we have accumulated considerable experience in studying the Omori law by theoretical and experimental methods. This paper summarizes the results of our study briefly. The main attention is focused on the phenomenological theory of aftershocks, the foundations of which were laid by Fusakichi Omori.

**Content**

1. Introduction
2. Elementary master equation
3. Logistic equation
4. Stochastic equation
5. Nonlinear diffusion equation

## 1. Introduction

Omori Law [1] describes the evolution of the aftershocks of a strong earthquake. Established at the end of the century before last, the law is characterized by the beauty of its form, quite definite clarity, as a result of which it still attracts considerable attention of the geophysical community (e.g., see [2–4]).

Initially, Omori law, which can be called hyperbolic, was formulated as follows:

$$n(t) = k/(c+t). \tag{1}$$

Here $n$ is the frequency of aftershocks [1]. Formula (1) is one-parameter, since the parameter $c$ is free and is completely determined by an arbitrary choice of the time origin. In contrast to the aspirations of Hirano [5] and Utsu [6], who introduced a two-parameter modification of formula (1) into widespread use, we came to the conviction that it is reasonable to put the differential equation of evolution into the basis of the phenomenological theory of aftershocks [7–9]. Guided by this consideration, in recent years we have accumulated considerable experience in the study of the evolution of aftershocks.

The purpose of this paper is to summarize our results. The main attention is focused on the phenomenological theory of aftershocks. The paper indicates successful examples of the use of theory in the analysis of experimental material. We also paid some attention to the presentation of our position on controversial issues of a methodological nature. For example, in the literature there are erroneous statements about the deep physical content of the parameter $c$ in formula (1).

Another misconception is associated with the idea that the two-parameter Hirano-Utsu formula is preferable to the one-parameter Omori formula. As an argument, it is argued that the presence of two parameters facilitates the approximation of observation data. On the contrary, we consider the presence of only one physical parameter in formula (1) as an important sign of the fundamental nature of the Omori law.

## 2. Elementary master equation

The differential approach to modeling aftershocks opens up a wide scope for searches. From the richest set of differential equations available here, we take the simplest implementation of our idea, namely the truncated Bernoulli equation

$$dn/dt + \sigma n^2 = 0, \qquad (2)$$

Here $\sigma$ is the so-called coefficient of deactivation of the earthquake source, "cooling down" after the main shock [7–9]. Elementary master equation (2) is useful in making the law of evolution simpler and easier to understand. It expresses the essence of the Omori hyperbolic law (1) that everyone understands. Moreover, it will serve as an initial basis for interesting generalizations (see below sections 3–5).

Equation (2) contains only one phenomenological parameter. It is easy to make sure that both formulations of the law, (1) and (2) are completely equivalent to each other for $\sigma = \text{const}$. However, firstly, in contrast to (1), formula (2) makes it possible to take into account the nonstationarity of the geological medium in the source, which undergoes a complex relaxation process after the discontinuity has formed during the main shock. The second advantage of formula (2) is no less important. We can seek and find natural generalizations of the differential law of evolution of aftershocks, which opens up new, sometimes unexpected approaches to processing and analyzing experimental data.

Concluding this section of the paper, let us show how easy it is to take into account nonstationarity when formulating the Omori law in the form of the evolution equation (2). For this, it is sufficient to assume that the deactivation factor depends on time. Let's rewrite Omori's law in the most compact form

$$dn/d\tau + n^2 = 0, \qquad (3)$$

where $\tau = \int_0^t \sigma(t')dt'$. The general solution to Eq. (3) is

$$n(\tau) = n_0 / (1 + n_0 \tau). \qquad (4)$$

It is seen that solution (4) retains the hyperbolic structure of the law, which was originally established thanks to Omori's discernment. The difference between (4) and (1) is only that time in the source, figuratively speaking, flows unevenly. For $\sigma = \text{const}$, (4) coincides with (1) up to notation. Thus, equation (2) and its solution (4) in a certain sense complete Omori's plan, as well as what Hirano and Utsu were striving for in their attempt to improve the law using a two-parameter modification of formula (1).

### 3. Logistic equation

Faraoni [10] considered the possibility of representing the Omori law, written in the form (2), as the Euler-Lagrange equation. The Lagrangian formulation of Omori law is interesting in many ways. In particular, it provides a basis for searching for possible generalizations of the law [11]. But the search for a suitable Lagrangian is not the only way to derive the evolution equation. One can proceed, for example, from a fairly general integro-differential equation

$$\frac{dn}{dt} = \int_0^\infty K(t-t') \cdot F[n(t')] \cdot dt'. \qquad (5)$$

If we put $F(n) = -\sigma n^2$, then when choosing the trivial kernel $K(t-t') = \delta(t-t')$ from (5) follows the Omori law in the form (2).

Derivation of (2) from (5) is useful in the sense that a natural generalization of the Omori law suggests to us. The need for generalization is dictated by the following consideration. It follows from (2) that $\lim n(t) = 0$ for $t \to 0$. Meanwhile, experience shows that the flow of aftershocks ends with a transition to a certain background seismicity of the source. It is desirable for us to take this circumstance into account by using minimal changes in the form of the classical Omori law. It turns out that for this it is enough to take into account the linear term in the formula $F(n)$:

$F(n) = \gamma n - \sigma n^2$. Here $\gamma$ is the second phenomenologically parameter of our theory. As a result, we get the master equation in the following form

$$dn/dt = n(\gamma - \sigma n). \qquad (6)$$

This is the logistic equation of Verhulst [12], well known in biology, chemistry and sociology. It turns out to be useful in the physics of earthquakes [9, 11].

We divide the family of solutions to logistic equation (6) into two classes. The first class includes growing, and the second, falling functions of time. The separation principle is easiest to show in the phase portrait shown in Figure 1, where we have used the dimensionless quantities

$$X = n/n_{max}, \quad P = (n_\infty/\gamma n_{max})dX/dT, \quad T = \gamma t, \qquad (7)$$

instead of the original quantities. Here $n_\infty = \gamma/\sigma$.

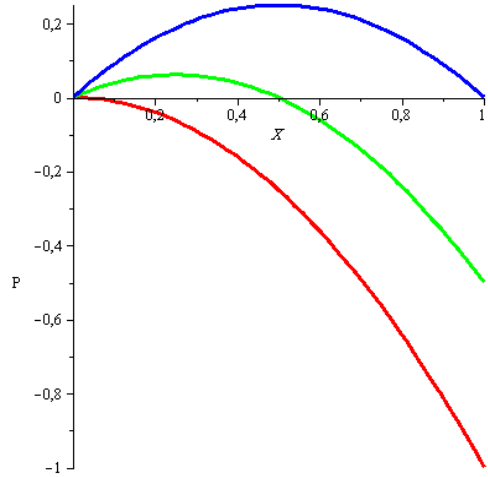

**Fig. 1.** Phase portrait on the phase plane of equation (6). The red, green and blue phase trajectories are plotted at $X_\infty = 0$, $0.5$ and $1$, respectively (see text).

Faraoni [10] proposed to introduce the phase plane of the dynamic system simulating the evolution of aftershocks according to the Omori law (2). The corresponding phase portrait is shown in Figure 1 with the red line. The point (0, 0) corresponds to the equilibrium state. The representative point moves from bottom to top along the phase trajectory with deceleration. The portrait consists of one phase trajectory that starts at point $(1,-1)$ and ends at point $(0,0)$.

However, we are interested in the family of phase trajectories for equation (6) constructed for different values of the parameter $X_\infty = n_\infty/n_{max}$. The red, green and blue trajectories in Figure 1 are plotted with $X_\infty$ values of $0$, $0.5$ and $1$, respectively. Equilibrium point $(0,0)$ is stable at $X_\infty = 0$ and unstable at $X_\infty > 0$. Equilibrium point $(X_\infty, 0)$ is stable at any values of parameter $X_\infty > 0$. It can be shown that the velocity of motion of the imaging point along the phase trajectory asymptotically tends to zero with approaching $(X_\infty, 0)$. The segment of the trajectory located above the horizontal axis corresponds to the Verhulst logistic curve, widely known in biology, chemistry, sociology and other sciences. The segment located below the horizontal axis corresponds to the evolution of aftershocks (Figure 2).

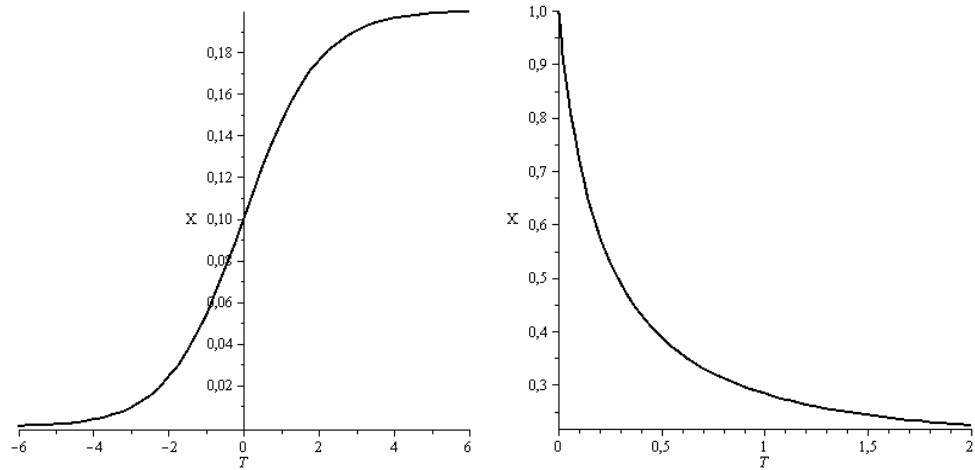

**Fig. 2.** Logistic curve (on left) and aftershocks curve (on right) at $X_\infty = 0.2$. Dimensionless time $T = \gamma t$ is plotted along the horizontal axis.

The choice between the logistic and aftershock branches is made when setting the Cauchy problem for equation (6). Evolution proceeds along the aftershock branch if the initial condition satisfies the inequality $n(0) = n_0 > n_\infty$, where $n_\infty = \gamma/\sigma$. Thus, in the physics of aftershocks, when setting the Cauchy problem, one should set the initial conditions under the additional constraint $n_0 > n_\infty$. Moreover, it is reasonable to use the strong inequality $n_0 \gg n_\infty = \gamma/\sigma$. Indeed, for $t \to \infty$, the frequency of aftershocks asymptotically approaches from above to the background (equilibrium) value $n_\infty$. Experience shows that as a rule $n_0 \gg n_\infty$ after a strong earthquake. The analysis of equation (6) under the condition $n_0 \gg n_\infty$ indicates that at the first stages of evolution, the frequency of aftershocks decreases with time in accordance with the classical Omori formula (1). Let's take a closer look at this important circumstance, since the existence of the aftershock branch is not so widely known.

The aftershock branch is entirely located above the saturation level $n_\infty$ and is a monotonically decreasing function of time. When $t \to +\infty$ it tends asymptotically from above the saturation level (see the right panel in Figure 2). When setting the Cauchy problem in the physics of aftershocks, the initial condition should be asked the restriction $n_0 \gg n_\infty$.

Well, let us show that the decrease in the frequency of aftershocks with time at the first stage of evolution occurs according to the Omori hyperbola (1). It is natural to call this stage of evolution the *Omori epoch*.

Let us introduce the notation

$$t_\infty = \frac{1}{\gamma} \ln\left(1 - \frac{n_\infty}{n_0}\right), \tag{8}$$

and write the solution of evolution equation (6) in the following form:

$$n(t) = n_\infty \left\{1 - \exp\left[\gamma(t_\infty - t)\right]\right\}^{-1}. \tag{9}$$

In the Omori epoch $t_\infty < t \ll 1/\gamma$ and respectively

$$n(t) = 1/\sigma(t - t_\infty). \tag{10}$$

Formula (10) coincides with the classical Omori formula (1) up to notation.

Observational experience indirectly testifies to the plausibility of our logistic model. It is known for example that over time the frequency $n$ tends not to zero, as follows from the Omori law, but to some equilibrium value $n_\infty$. Further, some combination of the logistic and aftershock branches makes it possible to propose a scenario for the occurrence of an earthquake swarm (see details in [11]).

## 4. Stochastic equation

Changing variables in a differential equation is often a powerful tool for finding solutions to it. We already know the solutions of the Omori equation (2) and the logistic equation (6). Nevertheless, we will still change the variable $n(t)$ in order to linearize both of these equations. This will make it easier for us to search for a stochastic generalization of the equation for the evolution of aftershocks.

The following replacement will help us transform nonlinear equations (2), (6) into linear ones [3]

$$n(t) \to g(t) = 1/n(t). \tag{11}$$

Omori equation (2) takes on an extremely simple form:

$$\dot{g} = \sigma. \qquad (12)$$

Here the dot above the symbol means time differentiation.

Logistic equation (6) becomes a first-order linear differential equation:

$$\dot{g} + \gamma g = \sigma. \qquad (13)$$

Generally speaking, this circumstance is quite interesting in itself, but we use it precisely to make the stochastic generalization of the evolution equation the simplest possible way. Namely, let's imagine that the deactivation coefficient experiences small fluctuations. This assumption is formalized as follows: $\sigma \to \sigma + \xi(t)$, where $\xi(t)$ is a random function of time, and $\max|\xi| \ll \sigma$. As a result, we have

$$\dot{g} + \gamma g = \sigma + \xi(t). \qquad (14)$$

Let us formally solve the equation (14)

$$g(t) = g_\infty + (g_0 - g_\infty)\exp(-\gamma t) + \int_0^t \xi(t_1)\exp\left[\gamma(t_1 - t)\right]dt_1. \qquad (15)$$

Here $g_\infty = \sigma/\gamma$, $g_0 = g(0)$.

Our second assumption is that $\xi(t)$ is the Langevin source, i.e. delta-correlated random function with zero mean

$$\overline{\xi(t)} = 0, \qquad \overline{\xi(t)\xi(t_1)} = N\delta(t - t_1), \qquad (16)$$

where the line from above means averaging.

Now equation (14) should be considered as the Langevin stochastic equation (e.g., see [13], where the Langevin equation is studied in detail). The phenomenological parameter $N$ is determined by the intensity of noise affecting our dynamic system.

## 5. Nonlinear diffusion equation

If we ask ourselves how to describe the evolution of aftershocks not only in time, but also in space-time, then this immediately puts us in a difficult position. On the one hand, a number of methods are known for modeling space-time distributions, but on the other hand, in our case, there is a strong limitation, which consists in the fact that when averaging over the epicentral zone, we want to obtain the Omori law in the form (2), or in the form (6). Fortunately for us, it turns out here that we can use the well-known Kolmogorov-Petrovsky-Piskunov equation (abbreviated KPP), which describes nonlinear diffusion [14]. It is convenient for us to represent it in the following form

$$\frac{\partial n}{\partial t} = n(\gamma - \sigma n) + D\frac{\partial^2 n}{\partial x^2}, \qquad (17)$$

where $n(x,t)$ is the spatio-temporal distribution of aftershocks, the $x$ axis is directed along the earth's surface (for simplicity, we limited ourselves to a one-dimensional model), and $D$ is a new phenomenological parameter (diffusion coefficient). At $D = 0$ (17) turns into the logistic equation of the evolution of aftershocks (6), and under the additional condition $\gamma = 0$ into Omori law (2).

It is useful to derive (17) from the integro-differential equation

$$\frac{\partial n}{\partial t} = \Phi(n) + \int_{-\infty}^{\infty} K(x-y) \cdot n(y,t) \cdot dy. \qquad (18)$$

This will make it possible to express the parameters $\gamma$ and $D$ in terms of the kernel $K(x-y)$. Indeed, suppose that $K(x-y) = K(y-x)$, i.e. the core is symmetrical. If $K \to 0$ for $|x-y| \to \infty$, then expanding $n(x-z,t)$ into a Taylor series in powers of $x$, we obtain

$$\frac{\partial n}{\partial t} = \gamma n + \Phi(n) + D\frac{\partial^2 n}{\partial x^2} + \ldots, \qquad (19)$$

where

$$\gamma = \int_{-\infty}^{\infty} K(z) \cdot dz, \quad D = \frac{1}{2}\int_{-\infty}^{\infty} z^2 K(z) \cdot dz, \qquad (20)$$

and $z = x - y$ (e.g., see [11, 15]). Let us $\Phi(n) = \sigma n^2$, and confine the first two terms in the series. In this approximation we obtain equation (17) from which after phenomenological reduction follows the Omori law in the form (2).

When constructing a phenomenological theory of aftershocks, there is one most important condition, only one, but absolutely necessary: the phenomenological coefficients, whatever meaning we put into them, we must be able to measure experimentally, since it is impossible to calculate them on the basis of a more fundamental theory - after all, such a theory we simply do not. In Section 6.1 we show how the deactivation factor we introduced in Section 2 is estimated. In the current section, we have significantly complicated the theory and introduced the diffusion coefficient $D$. We want to briefly describe the result of observing aftershocks, which actually led us to master equation (17), and then indicate how, at least in principle, the parameter $D$ can be estimated experimentally. (In this regard, it is appropriate to mention the recent interesting work [16]. It provides additional arguments in favor of the idea of the applicability of the KPP equation for modeling aftershocks.)

The study of aftershocks in space and time led us to the idea of using the KPP equation as the master equation. The main step forward in the study of the space-time distribution was the discovery that, apparently, at least some of the aftershocks tend to propagate like waves with a velocity much lower than the velocity of seismic waves [8, 17]. The rate of propagation varies widely from case to case. It is roughly a few kilometers per hour. This value is three orders of magnitude less than the velocities of elastic waves in the crust, which suggests the propagation of a nonlinear diffusion wave excited by the main shock.

The equation (17) has self-similar solutions in the form of a traveling wave $n(x,t) = n(x \pm Ut)$ [14, 18]. It is this circumstance that played a role in our choice of the KPP equation as the master equation. The estimation of the wave propagation velocity can be done by analyzing the dimensions of the coefficients of the master equation: $U \sim \sqrt{\gamma D}$. Knowing the propagation velocity $U$, and estimating the parameter $\gamma$ according to the formula $\gamma = n_\infty \sigma$, we can give an oriented estimate of the diffraction coefficient $D = U^2 / \gamma$.

## 6. Discussion

We have outlined a phenomenological basis, united by the general idea of a differential approach to describing and understanding the dynamics of aftershocks. Starting with an elementary nonlinear differential equation (2), completely equivalent to the Omori law in its classical expression (1) at $\sigma = \text{const}$, we tried to use minimal modifications in order to go first to the logistic equation (6), then to the stochastic equation (14) and, finally, to the nonlinear diffusion equation (17). Perhaps it would be useful to note that we explicitly used the methodological principle of Descartes, the essence of which is that one should go from simple to complex, using clear and precise modifications of the theoretical description of the problem under study.

Taken together, four phenomenological master equations, united by a common idea, make it possible to comprehend a fairly wide range of properties and patterns of aftershocks found experimentally. Moreover, phenomenological theory allows certain predictions to be made that can be verified experimentally. Let us illustrate what has been said with a number of examples.

### 6.1. Inverse problem

The inverse problem of the physics of earthquake source is to determine the phenomenological coefficients from the observation data of aftershocks. Omori law in the form (2) makes it possible to formulate and solve the problem of determining the deactivation coefficient $\sigma$ from the observation data of the frequency of aftershocks $n$. The auxiliary value $g$, which we introduce in Section 3, is conveniently written in the form of $g = (n_0 - n)/nn_0$. It is easy to see that

$$\sigma = \frac{d}{dt}\langle g \rangle. \tag{21}$$

Here, the angle brackets denote the regularization of the function $g(t)$ calculated from observation $n(t)$. Regularization is reduced in this case to the smoothing procedure.

Our experience indicates that the deactivation coefficient $\sigma(t)$ undergoes complex variations over time [8, 19]. However (and this seems to us extremely important) at the first stages of evolution $\sigma = \text{const}$. The time interval during which $\sigma = \text{const}$, we called the *Omori epoch*. In the Omori epoch, the classical Omori law is fulfilled (1), according to which the frequency of aftershocks hyperbolically decreases over time. In our experience, the duration of the Omori epoch varies from case to case from a few days to two or three months. We noticed a tendency for the duration of the Omori epoch to increase with the increase in the magnitude of the main shock.

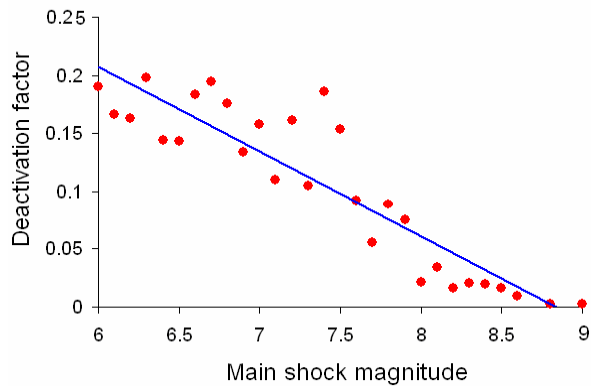

**Fig. 3.** Dependence of the deactivation factor of the earthquake source on the magnitude of the main shock.

An interesting prediction follows from the existence of the Omori epoch, which has been reliably confirmed by experience [20].

Namely, the question of dependence of the deactivation factor on the magnitude of main shock is analyzed theoretically and experimentally. A monotonic decrease in the deactivation factor with an increase in the magnitude of the main shock $M_0$ has been reliably established. Figure 3 shows the result of measurements of $\sigma$ at different values of $M_0$. To measure the deactivation factor, we used the technique developed during the compilation of the Atlas of Aftershocks [20]. We see that, on average, $\sigma$ decreases monotonically with the increase in $M_0$. The dependence $\sigma(M_0)$ is approximated by the formula

$$\sigma = A - BM_0, \qquad (22)$$

where $A = 0.64$, $B = 0.07$ with a sufficiently high coefficient of determination $R^2 = 0.82$. Thus, the theoretical inequality $d\sigma/dM_0 < 0$ is reliably confirmed by direct measurements. We have a wonderful harmony between theory and experiment.

It is quite clear that the question of how best to formulate the Omori law, in the form (1) or (2), could only be solved by observation and experience. Equation (2) turned out to be more effective, since it made it possible to introduce a simple and useful concept of deactivation of the source, to pose the inverse problem of the source, and to reveal the existence of the Omori epoch. In addition to this, we have shown that with the help of (2) one can make a meaningful statement about the deactivation coefficient and, moreover, check this statement experimentally.

Finally, the question of whether it is not better to use the two-parameter Hirkno-Utsu formula $n = k/(c+t)^p$ [2] for modeling aftershocks than the one-parameter formula (2) deserves discussion? We give preference to formula (2), since the inverse problem solved on its basis indicates the existence of the Omori epoch [8, 9]. The Hirano-Utsu formula is unacceptable, since it contradicts the existence of the Omori epoch at $p \neq 1$, and at $p = 1$ it coincides with the Omori formula (1).

Perhaps it would be appropriate to draw a distant historical analogy here. According to the law of gravitation, the interaction potential $\varphi \propto 1/r$ leads to the ellipticity of the planetary orbit. The deviation from ellipticity, for example, of the orbit of Mercury, could in no way serve as a reason for choosing the interaction, say, in form $\varphi \propto 1/r^p$. Other understanding of the deviation of orbit from strictly elliptical had to be looked for, and it was foun within the framework of general relativity. Perhaps, finding themselves in a similar situation, Hirano and Utsu should not have immediately abandoned the excellent Omori law, but should have looked for other explanations for the deviation of the real flow of aftershocks from strict hyperbolicity.

### 6.2. Triggers

Deviations from the classical Omori law (1) are caused not only by the nonstationarity of the parameters of the geological environment in the source, which we have expressed in the form of a possible dependence of the deactivation factor on time. Deviations can occur under the influence of so-called triggers, i.e. relatively small disturbances of geophysical fields of endogenous or exogenous origin.

We point here to two endogenous triggers that we have recently discovered. Both are aroused on the main shock. One of them has the form of a round-the-world seismic echo, and the second represents free elastic oscillations of the Earth as a whole, excited by the main shock. We have described both triggers in detail in a number of papers, so we will restrict ourselves here to references [21–25].

Let's dwell on exogenous triggers in more detail. For a long time the cosmic effects on seismicity have been widely discussed but there is still no agreement in the geophysical community on the effectiveness of such impacts. The controversy about the influence of geomagnetic storms on the global activity of earthquakes arises especially often (see, for example, [26–29]). The question is really difficult. On the one hand, observations indicate a correlation between seismicity and geomagnetic storms and a complex of electromagnetic phenomena associated with them (see papers [30–37] and the literature cited therein). On the other hand, the mechanism of the impact of geomagnetic storms on rocks, leading to modulation of seismicity, is not entirely clear. In this regard, the idea of the magnetoplasticity of rocks [31, 32] seems to us very encouraging, but a discussion of this deep idea would lead us far aside.

A wide class of exogenous triggers of anthropogenic origin is known. We will restrict ourselves here to an indication of the weekend effect discovered in [38], and the so-called Big Ben effect, or the effect of hour markers [39. 40]. Both effects pose a difficult question for the researcher about the global impact of the industrial activity of mankind on the lithosphere.

### 6.3. Triads

Foreshocks, main shock and aftershocks are perceived by us as a ready-made and complete set. We propose to call the peculiar trinity of foreshocks, main shock and aftershocks a *triad*. The magnitude of the main shock is always greater than the maximum magnitudes of foreshocks and aftershocks. The classical triad satisfies the inequalities

$$M_- < M_+, \qquad (23)$$

and

$$N_- < N_+. \qquad (24)$$

Here $M_-$ ($M_+$) and $N_-$ ($N_+$) are the maximum magnitude and the number of foreshocks (aftershocks), respectively.

We want to draw attention to the existence of anomalous triads for which inequalities (23), (24) do not hold [41]. These are the so-called mirror triads, for which $M_- > M_+$, $N_- > N_+$, and symmetric triads, for which $N_- = N_+$. Abnormal triads are few in number. According to our calculations, the number of anomalous triads is about an order of magnitude less than the classical triads, but they are undoubtedly of exceptional interest in our attempts to understand the dynamics of earthquakes.

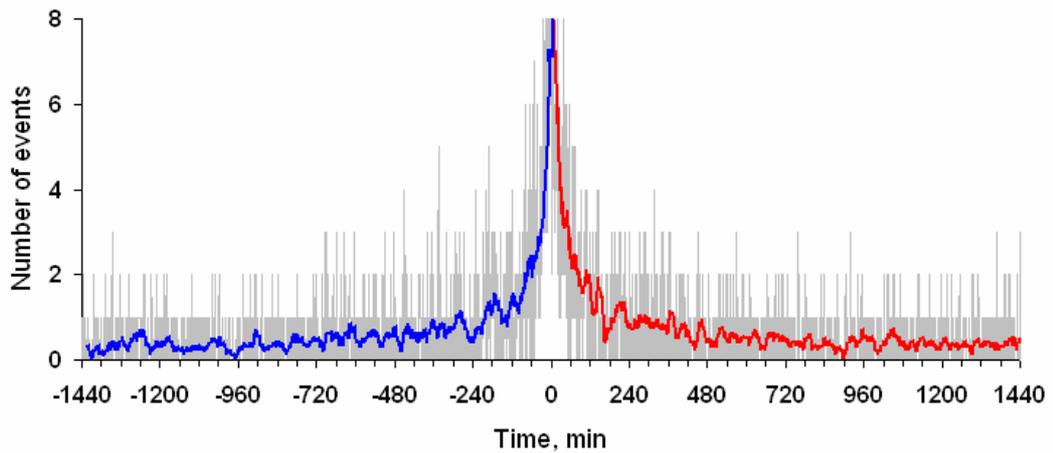

**Fig. 4.** Symmetrical triads. Zero moment of time corresponds to the moment of the main shock.

Figure 4 illustrates the time distribution of symmetric triads. We used data on earthquakes that occurred on Earth from 1973 to 2019 and were registered in the world catalog of earthquakes USGS/NEIC (https://earthquake.usgs.gov). Note that the mirror analogue of the Omori law is fulfilled for foreshocks. This property is possessed not only by symmetrical, but also by mirror triads.

So, we know that in the classical triad, the number of foreshocks is less than the number of aftershocks. In many cases, foreshocks are completely absent even before very strong earthquakes. We asked the question: Are there earthquakes that are strong enough (say, with a magnitude of $M_0 \geq 6$), neither before nor after which there are neither foreshocks nor aftershocks? The search result was amazing. We have discovered a wide variety of this kind of earthquake and named it *Grande terremoto solitario*, or GTS for short. The number of GTS is approximately equal to the number of classical triads.

GTS arise spontaneously under very calm seismic conditions, and are not accompanied by aftershocks. This suggests an analogy between the GTS and the so-called "Rogue waves" (or "Freak waves") – isolated giant waves that occasionally emerge on a relatively quiet ocean surface (e.g., see [42]). This analogy may prove to be quite profound, since the spontaneous occurrence of pulses having anomalously high amplitudes is a common property of the nonlinear evolution of dynamic systems [43].

The mirror triads, these ghosts of the classical triads, are not only curious in themselves, but can most decisively influence our understanding of the alternative possibilities of the dynamics of lithosphere, leading to catastrophic earthquakes. In particular, we face a fundamental problem, the essence of which is to find the physical and geotectonic reasons for the apparent predominance of truncated classical triads in the seismic activity of the Earth.

Concluding the discussion, we want to make a judgment, perhaps controversial, that Omori's law, like no other, gives us the opportunity to realize the uncertainty, incompleteness and, in a certain sense, immaturity of the physics of earthquakes. So far, it is still possible and in reality there is a wide range of opinions on the unresolved issues of the Earth's seismicity. So far, everyone, in accordance with their idea of the nature of cognition, can see in the Omori law either an empirical formula convenient for approximating observations, or an indication of the deep physical meaning of hyperbolicity in the frequency of aftershocks after a strong earthquake.

# 7. Conclusion

We will deviate from tradition and, instead of simply listing the results of the work (which is still far from complete) we will briefly present some prospect of the research that we propose to carry out in the near future. The study will be devoted to the geometrodynamics of a tectonic earthquake source. We proceed from the fact that in the study of earthquakes some geometrically visual representations and considerations are necessary, and that analytics alone is insufficient. Representing the source as the interior of the convex hull of a point manifold of aftershocks, we have outlined a program whose purpose is to reduce the mosaic of very complex, intricate realities of the evolving source to geometric objects. An interesting object is the space-time trajectory of the shell's center of gravity. The curvature of the envelope surface of the epicenters in dynamics and other geometrically visual images can also turn out to be quite interesting. As conceived, the geometrodynamics will become a source of new ideas for the development of the phenomenological theory of earthquakes.

However, let us return to the work presented by us and try to make some summary of all the results. The general conclusion is as follows: the methodological approach based on differential equations of evolution opens up new possibilities for the analysis of experimental material. The phenomenological equations of evolution proposed by us allow to pose inverse problems of the source physics and make it possible to formulate unexpected questions regarding the dynamics of earthquakes. The phenomenological theory, a sketch of which we have given here, not only enriches the system of ideas about the source, but, we hope, indicates the possibility of searching for approaches to solving problems of a fundamental nature.

We are fully aware of the fact that neither the totality of facts, even if it is represented by a set of empirical formulas, nor a logically consistent phenomenological theory, by itself, does not lead us to a deep penetration into the essence of earthquakes. A deep understanding would be much more facilitated by a theory based on the fundamental laws of physics, taking into account the characteristics of the geological environment. Theoretical constructions of this kind are known, but for a completely understandable reason they refer only to individual aspects of the phenomenon, and not to the phenomenon as a whole. Under these conditions, it is reasonable and natural to consistently continue the development of the phenomenological theory of earthquakes, the foundations of which were laid by Fusakichi Omori.

*Acknowledgments*. We express our deep gratitude to A.L. Buchachenko, F.Z. Feygin, N.A Kurazhkovskaya and A.S. Potapov for interest in this work. The authors thank the staff U.S. Geological Survey for providing the catalogues of earthquakes. The work was carried out according to the plan of state assignments of IPhE RAS.